%% file: ExactSol_MasterEq_II.tex
\documentclass[12pt,a4paper]{article}
\usepackage{amsmath}
\usepackage{latexsym}
\makeatletter
\def\rddots{\mathinner{\mkern1mu\raise\p@%
    \vbox{\kern7\p@\hbox{.}}\mkern2mu%
    \raise4\p@\hbox{.}\mkern2mu\raise7\p@\hbox{.}\mkern1mu}}
\makeatother
\newcommand{\ket}[1]{{\vert{#1}\rangle}}
\newcommand{\bra}[1]{{\langle{#1}\vert}}
\newcommand{\kett}[1]{{\vert{#1}\rangle\rangle}}
\newcommand{\braa}[1]{{\langle\langle{#1}\vert}}

\newcommand{\fukuso}{{\mathbf C}}

\newcommand{\tr}{{\rm tr}}

\begin{document}

\title{\sl Exact Solution of a Master Equation Applied to 
the Two Level System of an Atom}
\author{
  Kazuyuki FUJII
  \thanks{E-mail address : fujii@yokohama-cu.ac.jp }\\
  International College of Arts and Sciences\\
  Yokohama City University\\
  Yokohama, 236--0027\\
  Japan
  }
\date{}
\maketitle
\begin{abstract}
  In this paper we discuss a master equation applied to 
the two level system of an atom and derive an exact 
solution to it in an abstract manner. We also present 
a problem and a conjecture based on the three level system.

Our results may give a small hint to understand the huge 
transition from Quantum World to Classical World.

To the best of our knowledge this is the finest method 
up to the present.
\end{abstract}
\vspace{5mm}\noindent

\vspace{5mm}\noindent

\section{Introduction}
The target of this paper is to study and solve the time evolution 
of a quantum state (which is a superposition of two physical states) 
under {\bf decoherence}.

In order to set the stage and to introduce proper notation, 
let us start with a system of principles of Quantum 
Mechanics (QM in the following for simplicity). See for example \cite{PD}, 
\cite{HG}, \cite{AP} and \cite{AH}. That is,

\vspace{5mm}\noindent
\begin{Large}
{\bf System of Principles of QM}
\end{Large}

\vspace{3mm}\noindent
1.\ {\bf Superposition Principle}\\
If $\ket{a}$ and $\ket{b}$ are physical states then their 
superposition $\alpha\ket{a}+\beta\ket{b}$ is also 
a physical state where $\alpha$ and $\beta$ are 
complex numbers.

\vspace{3mm}\noindent
2.\ {\bf Schr\"{o}dinger Equation and Evolution}\\
Time evolution of a physical state proceeds like
\[
\ket{\Psi}\ \longrightarrow\ U(t)\ket{\Psi}
\]
where $U(t)$ is the unitary evolution operator 
($U^{\dagger}(t)U(t)=U(t)U^{\dagger}(t)={\bf 1}$ and $U(0)={\bf 1}$) 
determined by a Schr\"{o}dinger Equation.

\vspace{3mm}\noindent
3.\ {\bf Copenhagen Interpretation}\footnote{There are some 
researchers who are against this terminology, see for 
example \cite{AH}. However, I don't agree with them 
because the terminology is nowadays very popular in the 
world}\\
Let $a$ and $b$ be the eigenvalues of an observable $Q$, and 
$\ket{a}$ and $\ket{b}$ be the normalized eigenstates corresponding to 
$a$ and $b$. When a state is a superposition $\alpha\ket{a}+\beta\ket{b}$ 
and we observe the observable $Q$ the state collapses like
\[
\alpha\ket{a}+\beta\ket{b}\ \longrightarrow \ \ket{a}
\quad\mbox{or}\quad
\alpha\ket{a}+\beta\ket{b}\ \longrightarrow \ \ket{b}
\]
where their collapsing probabilities are $|\alpha|^{2}$ and 
$|\beta|^{2}$ respectively ($|\alpha|^{2}+|\beta|^{2}=1$).

This is called the collapse of the wave function and 
the probabilistic interpretation.

\vspace{3mm}\noindent
4.\ {\bf Many Particle State and Tensor Product}\\
A multiparticle state can be constructed by the superposition of 
the Knonecker products of one particle states, which are called 
the tensor products. For example, 
\[
\alpha|{a}\rangle\otimes|{a}\rangle +\beta|{b}\rangle\otimes|{b}\rangle
\equiv 
\alpha|{a,a}\rangle +\beta|{b,b}\rangle
\]
is a two particle state.

\vspace{5mm}
Here is an important comment. Beginners of QM  might think 
that a quantum state created by an experiment would undergo 
the unitary time evolution (U) forever. 

This is nothing but an illusion because the quantum state is in 
an environment (a kind of heat bath) 
and the interaction with it will disturb the quantum state. 
For example, readers should imagine an oscillator on the desk. 

In order to understand QM deeply readers should take 
decoherence (: interaction with environment) into consideration 
correctly. For this topic see for example \cite{WZ}.

\section{Two Level System and Master Equation}
In this section we discuss a master equation applied to 
the two level system of an atom and solve the equation 
exactly under certain conditions.

For the discussion of the two level system of an atom 
let us prepare some notations from Quantum Optics. 
See for example \cite{WS}, \cite{Five}.

For the system the target space is  
${\bf C}^{2}=\mbox{Vect}_{{\bf C}}(|{0}\rangle, |{1}\rangle)$ 
with bases
\[
|{0}\rangle=
\left(
\begin{array}{c}
1 \\
0
\end{array}
\right),\quad
|{1}\rangle=
\left(
\begin{array}{c}
0 \\
1
\end{array}
\right).
\]
Then Pauli matrices $\{\sigma_{1},\ \sigma_{2},\ \sigma_{3}\}$ 
with the identity $1_{2}$
\[
\sigma_{1}=
\left(
\begin{array}{cc}
0 & 1 \\
1 & 0
\end{array}
\right),\quad
\sigma_{2}=
\left(
\begin{array}{cc}
0 & -i \\
i  & 0
\end{array}
\right),\quad
\sigma_{3}=
\left(
\begin{array}{cc}
1 & 0  \\
0 & -1
\end{array}
\right),\quad
1_{2}=
\left(
\begin{array}{cc}
1 & 0 \\
0 & 1
\end{array}
\right)
\]
act on the space. By setting
\[
\sigma_{+}\equiv \frac{1}{2}(\sigma_{1}+i\sigma_{2})=
\left(
\begin{array}{cc}
0 & 1 \\
0 & 0
\end{array}
\right),\quad
\sigma_{-}\equiv \frac{1}{2}(\sigma_{1}-i\sigma_{2})=
\left(
\begin{array}{cc}
0 & 0 \\
1 & 0
\end{array}
\right)
\]
it is easy to see
\[
\sigma_{+}\sigma_{-}=
\left(
\begin{array}{cc}
1 & 0 \\
0 & 0
\end{array}
\right),\quad
\sigma_{-}\sigma_{+}=
\left(
\begin{array}{cc}
0 & 0 \\
0 & 1
\end{array}
\right).
\]

For the initial time $t=0$ we may assume that the Hamiltonian 
(of the atom) is of a diagonal form
\begin{equation}
\label{eq:diagonal matrix}
H_{0}=
\left(
\begin{array}{cc}
E_{0} & 0  \\
0 & E_{1}
\end{array}
\right)
\end{equation}
where $E_{0}$ and $E_{1}$ are the two eigenvalues 
($E_{0}<E_{1}$ for simplicity) of the atom. It is easy to see
\[
H_{0}\ket{0}=E_{0}\ket{0},\quad H_{0}\ket{1}=E_{1}\ket{1}.
\]

For $t>0$ we consider an interaction of the atom 
with some laser field. Then the interaction term is 
included as the non-diagonal terms of the Hamiltonian
\begin{equation}
\label{eq:interaction matrix}
H=
\left(
\begin{array}{cc}
E_{0} & \gamma         \\
\bar{\gamma} & E_{1}
\end{array}
\right).
\end{equation}
Here we assume for simplicity that $\gamma$ is a complex constant. 

First, let us calculate the eigenvalues of the interacting 
Hamiltonian (\ref{eq:interaction matrix}) :
\begin{eqnarray*}
&&
0=|\lambda 1_{2}-H|
=
\left|
\begin{array}{cc}
\lambda -E_{0} & -\gamma         \\
-\bar{\gamma} & \lambda -E_{1}
\end{array}
\right|
=\lambda^{2}-(E_{0}+E_{1})\lambda +E_{0}E_{1}-|\gamma|^{2} \\
&&
\Longrightarrow\ 
\lambda_{\pm}=\frac{E_{0}+E_{1}\pm \sqrt{(E_{1}-E_{0})^{2}+4|\gamma|^{2}}}{2}.
\end{eqnarray*}
Note the order
\[
\lambda_{+}>E_{1}>E_{0}>\lambda_{-}.
\]

Next, the eigenvector of $\lambda_{-}$ is given by
\begin{eqnarray*}
\ket{\lambda_{-}}
&=&
\frac{|\gamma|}{\sqrt{|\gamma|^{2}+(E_{0}-\lambda_{-})^{2}}}
\left(
\begin{array}{c}
1 \\
-\frac{E_{0}-\lambda_{-}}{\gamma}
\end{array}
\right) \\
&=&
\frac{|\gamma|}{\sqrt{|\gamma|^{2}+(E_{0}-\lambda_{-})^{2}}}\ket{0}-
\frac{|\gamma|}{\gamma}
\frac{E_{0}-\lambda_{-}}{\sqrt{|\gamma|^{2}+(E_{0}-\lambda_{-})^{2}}}\ket{1}
\end{eqnarray*}
(we omit the details of $\ket{\lambda_{+}}$).  

\noindent
This state (having the eigenvalue $\lambda_{-}$) is just 
a superposition of $\ket{0}$ and $\ket{1}$. 
This example shows that a superposition of quantum states 
can lower the energy level. Maskawa says in \cite{TM} 
that this phenomenon is the essence of superposition in QM.

See the following figure : 
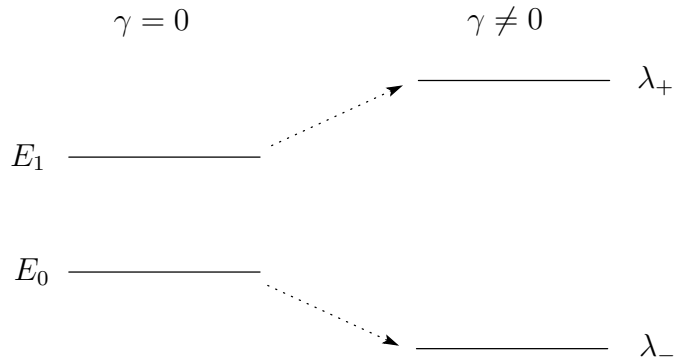
\vspace{8mm}
\begin{center}
\input{eigenvalues.tex}
\end{center}
\vspace{5mm}

From now on we study the time evolution of (\ref{eq:interaction matrix}) 
including decoherence interactions. 
To treat the decoherence phenomena in a correct manner 
it is important to adopt the density matrix formulation \footnote{
This point is a bit difficult to understand for beginners} 
instead of the pure state formulation discussed on far. 
The general definition of density matrix $\rho$ is 
\[
\rho^{\dagger}=\rho \ \   \mbox{and} \ \  \tr{\rho}=1,
\]
so we can write $\rho=\rho(t)$ as
\begin{equation}
\label{eq:density matrix}
\rho=
\left(
\begin{array}{cc}
a         & b  \\
\bar{b} & d
\end{array}
\right)
\quad (a=\bar{a},\ d=\bar{d},\ a+d=1).
\end{equation}
Here we have suppressed the $t$ dependence of the 
components like $a=a(t)$, etc for simplicity.

The general form of the master equation (\cite{Lind}, \cite{GKS} 
or \cite{BP}) is well--known to be
\begin{equation}
\label{eq:master equation 1}
\frac{d}{dt}\rho=-i[H, \rho]+D\rho \quad (\Leftarrow \hbar=1\ 
\mbox{for simplicity})
\end{equation}
where
\[
D\rho=
\mu\left(\sigma_{-}\rho\sigma_{+}-\frac{1}{2}\sigma_{+}\sigma_{-}\rho-\frac{1}{2}\rho\sigma_{+}\sigma_{-}\right)
+
\nu\left(\sigma_{+}\rho\sigma_{-}-\frac{1}{2}\sigma_{-}\sigma_{+}\rho-\frac{1}{2}\rho\sigma_{-}\sigma_{+}\right)
\]
and $\mu$ and $\nu$ are positive constants 
($\mu,\ \nu>0$) representing phenomenologically the feeble 
interactions with the environment. 
Note that $\mu$ and $\nu$ are determined by models.

We must solve the equation (\ref{eq:master equation 1}). 
By use of the transformation
\[
\rho=
\left(
\begin{array}{cc}
a         & b  \\
\bar{b} & d
\end{array}
\right)
\ \longrightarrow \ 
\hat{\rho}=
\left(
\begin{array}{c}
a  \\
b  \\
\bar{b} \\
d
\end{array}
\right)
\]
the master equation can be rewritten as
\begin{equation}
\label{eq:master equation 2}
\frac{d}{dt}
\left(
\begin{array}{c}
a \\
b \\
\bar{b} \\
d
\end{array}
\right)
=
\left(
\begin{array}{cccc}
-\mu & i\bar{\gamma} & -i\gamma & \nu                                         \\
i\gamma & i(E_{1}-E_{0})-\frac{\mu+\nu}{2} & 0 & -i\gamma                  \\
-i\bar{\gamma} & 0 & -i(E_{1}-E_{0})-\frac{\mu+\nu}{2} & i\bar{\gamma} \\
\mu & -i\bar{\gamma} & i\gamma & -\nu 
\end{array}
\right)
\left(
\begin{array}{c}
a \\
b \\
\bar{b} \\
d
\end{array}
\right).
\end{equation}
The derivation is left to readers. For example, refer to \cite{KF2}.

First, we must look for eigenvalues of the matrix $W$
\begin{equation}
\label{eq:W}
W=
\left(
\begin{array}{cccc}
-\mu & i\bar{\gamma} & -i\gamma & \nu                                         \\
i\gamma & i(E_{1}-E_{0})-\frac{\mu+\nu}{2} & 0 & -i\gamma                  \\
-i\bar{\gamma} & 0 & -i(E_{1}-E_{0})-\frac{\mu+\nu}{2} & i\bar{\gamma} \\
\mu & -i\bar{\gamma} & i\gamma & -\nu 
\end{array}
\right),
\end{equation}
which is very hard. Since
\begin{eqnarray*}
0&=&|\lambda 1_{4}-W| \\
&=&
\left|
\begin{array}{cccc}
\lambda+\mu & -i\bar{\gamma} & i\gamma & -\nu                                        \\
-i\gamma & \lambda-i(E_{1}-E_{0})+\frac{\mu+\nu}{2} & 0 & i\gamma                 \\
i\bar{\gamma} & 0 & \lambda+i(E_{1}-E_{0})+\frac{\mu+\nu}{2} & -i\bar{\gamma}  \\
-\mu & i\bar{\gamma} & -i\gamma & \lambda+\nu 
\end{array}
\right| \\
&=& \cdots \\
&=&\lambda
\left|
\begin{array}{cccc}
1 & 0 & 0 & 0                                                                                             \\
-i\gamma & \lambda-i(E_{1}-E_{0})+\frac{\mu+\nu}{2} & 0 & 2i\gamma                 \\
i\bar{\gamma} & 0 & \lambda+i(E_{1}-E_{0})+\frac{\mu+\nu}{2} & -2i\bar{\gamma}  \\
-\mu & i\bar{\gamma} & -i\gamma & \lambda+\mu+\nu 
\end{array}
\right| \\
&=&\lambda
\left|
\begin{array}{ccc}
\lambda-i(E_{1}-E_{0})+\frac{\mu+\nu}{2} & 0 & 2i\gamma           \\
0 & \lambda+i(E_{1}-E_{0})+\frac{\mu+\nu}{2} & -2i\bar{\gamma}  \\
i\bar{\gamma} & -i\gamma & \lambda+\mu+\nu 
\end{array}
\right| \\
&=&\lambda
\left[
\left\{\left(\lambda+\frac{\mu+\nu}{2}\right)^{2}+(E_{1}-E_{0})^{2}\right\}(\lambda+\mu+\nu)+
2|\gamma|^{2}(2\lambda+\mu+\nu)
\right]
\end{eqnarray*}
we obtain one trivial root $\lambda=0$ and a cubic equation
\[
\left\{\left(\lambda+\frac{\mu+\nu}{2}\right)^{2}+(E_{1}-E_{0})^{2}\right\}(\lambda+\mu+\nu)+
2|\gamma|^{2}(2\lambda+\mu+\nu)=0.
\]
Let us transform this. By setting
\[
\Lambda=\lambda+\frac{\mu+\nu}{2}\ \Longrightarrow\ 
\lambda =\Lambda-\frac{\mu+\nu}{2}
\]
the cubic equation becomes
\begin{equation}
\label{eq:Cubic}
\Lambda^{3}+\frac{\mu+\nu}{2}\Lambda^{2}+
\{(E_{1}-E_{0})^{2}+4|\gamma|^{2}\}\Lambda+(E_{1}-E_{0})^{2}\frac{\mu+\nu}{2}=0.
\end{equation}

Since the equation is cubic we can solve it  by use of 
the Cardano formula formally. See for example \cite{KF4}. 
However, the formula does not suit our purpose well.

Here we set
\[
f(\Lambda)=
\Lambda^{3}+\frac{\mu+\nu}{2}\Lambda^{2}+
\{(E_{1}-E_{0})^{2}+4|\gamma|^{2}\}\Lambda+(E_{1}-E_{0})^{2}\frac{\mu+\nu}{2}
\]
and treat its roots in an abstract way. Note that $f(\Lambda)>0$ 
for $\Lambda\geq 0$ because all coefficients are positive. 
Since
\[
f(0)=(E_{1}-E_{0})^{2}\frac{\mu+\nu}{2}>0
\quad \mbox{and}\quad
f(-\frac{\mu+\nu}{2})=-2|\gamma|^{2}(\mu+\nu)<0
\]
there is (at least) one root $-\frac{\mu+\nu}{2}<\Lambda_{0}<0$ 
satisfying $f(\Lambda_{0})=0$. By denoting
\[
f(\Lambda)=\Lambda^{3}+a\Lambda^{2}+b\Lambda+ c
\]
for simplicity we have a decomposition
\[
f(\Lambda)=(\Lambda-\Lambda_{0})
(\Lambda^{2}+(\Lambda_{0}+a)\Lambda +(\Lambda_{0}^{2}+a\Lambda_{0}+b))=0.
\]
From this we obtain other two roots
\[
\Lambda_{\pm}=\frac{-(\Lambda_{0}+a)\pm 
\sqrt{(\Lambda_{0}+a)^{2}-4(\Lambda_{0}^{2}+a\Lambda_{0}+b)}}{2}.
\]
Note that $\Lambda_{0}+a=\Lambda_{0}+\frac{\mu+\nu}{2}>0$.

If $\Lambda_{0}^{2}+a\Lambda_{0}+b<0$ then $\Lambda_{+}>0$, 
which is a contradiction. Therefore, $\Lambda_{0}^{2}+a\Lambda_{0}+b>0$.

As a result,
\[
\Lambda_{-}<\Lambda_{+}<0
\]
if $(\Lambda_{0}+a)^{2}-4(\Lambda_{0}^{2}+a\Lambda_{0}+b)>0$ (real 
roots) and
\[
\Lambda_{\pm}=\frac{-(\Lambda_{0}+a)\pm i
\sqrt{4(\Lambda_{0}^{2}+a\Lambda_{0}+b)-(\Lambda_{0}+a)^{2}}}{2}
\]
if $(\Lambda_{0}+a)^{2}-4(\Lambda_{0}^{2}+a\Lambda_{0}+b)<0$ 
(complex conjugate roots). 
In this case, the real part is negative
\[
\mbox{Re}\ \Lambda_{\pm}=-\frac{\Lambda_{0}+a}{2}<0.
\]

The solutions of the characteristic polynomial of $W$ 
(= $|\lambda 1_{4}-W|$) are
\begin{equation}
\label{eq:solutions}
\lambda_{1}=0,\quad
\lambda_{2}=\Lambda_{0}-\frac{\mu+\nu}{2},\quad
\lambda_{3}=\Lambda_{+}-\frac{\mu+\nu}{2},\quad
\lambda_{4}=\Lambda_{-}-\frac{\mu+\nu}{2}
\end{equation}
and
\begin{equation}
\label{eq:basic properties}
\lambda_{2}<0,\quad \lambda_{3}<0,\quad \lambda_{4}<0
\quad
\mbox{or}
\quad
\lambda_{2}<0,\quad \mbox{Re}\lambda_{3}<0,\quad \mbox{Re}\lambda_{4}<0
\end{equation}
under the conditions stated above.

Next, we look for the eigenvectors corresponding to 
the eigenvalues. For the purpose let us prepare some notations. 
We use the convension that 
the ket vecor $\ket{\lambda}$ is normalized, while the round 
ket vector $|\lambda)$ is not normalized 
(that is, $\langle{\lambda}|{\lambda}\rangle=1$ and 
$({\lambda}|{\lambda})\ne 1$).

It is easy to obtain the eigenvectors of $W^{T}$ 
rather than those of $W$ as shown in the following. 
Namely,
\begin{equation}
\label{eq:W^{T}}
W^{T}=
\left(
\begin{array}{cccc}
-\mu & i\gamma & -i\bar{\gamma} & \mu                                        \\
i\bar{\gamma} & i(E_{1}-E_{0})-\frac{\mu+\nu}{2} & 0 & -i\bar{\gamma}  \\
-i\gamma & 0 & -i(E_{1}-E_{0})-\frac{\mu+\nu}{2} & i\gamma                \\
\nu & -i\gamma & i\bar{\gamma} & -\nu
\end{array}
\right).
\end{equation}
Of course, $W$ and $W^{T}$ share the same eigenvalues. 
Let us list the eigenvectors of $W^{T}$ :
\[
|\lambda_{1})=
\left(
\begin{array}{c}
1 \\
0 \\
0 \\
1
\end{array}
\right)
\]
and we set
\[
|\lambda_{2})=
\left(
\begin{array}{c}
x_{2} \\
y_{2} \\
z_{2} \\
w_{2}
\end{array}
\right),
\quad
|\lambda_{3})=
\left(
\begin{array}{c}
x_{3} \\
y_{3} \\
z_{3} \\
1
\end{array}
\right),
\quad
|\lambda_{4})=
\left(
\begin{array}{c}
x_{4} \\
y_{4} \\
z_{4} \\
1
\end{array}
\right).
\]
See the next section why we make such a choice.

\vspace{3mm}
{\bf Note}. Let us show how to construct an eigenvector 
$|\lambda)$ from the eigenvalue $\lambda$. In order to 
avoid complicated expressions (equations) 
we restrict to the case of $n=3$. 
That is, the equation is
\[
\left(
\begin{array}{ccc}
a_{1} & b_{1} & c_{1} \\
a_{2} & b_{2} & c_{2} \\
a_{3} & b_{3} & c_{3}
\end{array}
\right)
\left(
\begin{array}{c}
x \\
y \\
z
\end{array}
\right)
=
\lambda
\left(
\begin{array}{c}
x \\
y \\
z
\end{array}
\right).
\]
From the first and second rows we have
\[
\left(
\begin{array}{ccc}
a_{1} & b_{1} \\
a_{2} & b_{2}
\end{array}
\right)
\left(
\begin{array}{c}
x \\
y
\end{array}
\right)
+
\left(
\begin{array}{c}
c_{1}z \\
c_{2}z
\end{array}
\right)
=
\lambda
\left(
\begin{array}{c}
x \\
y
\end{array}
\right)
\]
or
\[
\left(
\begin{array}{ccc}
\lambda-a_{1} & -b_{1} \\
-a_{2} & \lambda-b_{2}
\end{array}
\right)
\left(
\begin{array}{c}
x \\
y
\end{array}
\right)
=
z
\left(
\begin{array}{c}
c_{1} \\
c_{2}
\end{array}
\right).
\]
If we asuume that the determinant is non-zero
\[
\left|
\begin{array}{ccc}
\lambda-a_{1} & -b_{1} \\
-a_{2} & \lambda-b_{2}
\end{array}
\right|
=(\lambda-a_{1})(\lambda-b_{2})-a_{2}b_{1}\ne 0
\]
we have
\begin{eqnarray*}
\left(
\begin{array}{c}
x \\
y
\end{array}
\right)
&=&
z
\left(
\begin{array}{ccc}
\lambda-a_{1} & -b_{1} \\
-a_{2} & \lambda-b_{2}
\end{array}
\right)^{-1}
\left(
\begin{array}{c}
c_{1} \\
c_{2}
\end{array}
\right) \\
&=&
\frac{z}{(\lambda-a_{1})(\lambda-b_{2})-a_{2}b_{1}}
\left(
\begin{array}{c}
(\lambda-b_{2})c_{1}+b_{1}c_{2} \\
a_{2}c_{1}+(\lambda-a_{1})c_{2}
\end{array}
\right).
\end{eqnarray*}
Therefore, we obtain
\[
\left(
\begin{array}{c}
x \\
y \\
z
\end{array}
\right)
=
\frac{z}{(\lambda-a_{1})(\lambda-b_{2})-a_{2}b_{1}}
\left(
\begin{array}{c}
(\lambda-b_{2})c_{1}+b_{1}c_{2} \\
a_{2}c_{1}+(\lambda-a_{1})c_{2} \\
(\lambda-a_{1})(\lambda-b_{2})-a_{2}b_{1}
\end{array}
\right).
\]
As a result, the eigenvector $|\lambda)$ is given by
\[
|\lambda)
=
\left(
\begin{array}{c}
(\lambda-b_{2})c_{1}+b_{1}c_{2} \\
a_{2}c_{1}+(\lambda-a_{1})c_{2} \\
(\lambda-a_{1})(\lambda-b_{2})-a_{2}b_{1}
\end{array}
\right).
\]
If the determinant above is zero then we have only to 
apply the same procedure to other two rows. 

For the readers let us give one exercise :
\[
A=
\left(
\begin{array}{ccc}
2 & -1 & 0  \\
-1 & 2 & -1 \\
0 & -1 & 2
\end{array}
\right).
\]

\vspace{3mm}
If we set
\begin{equation}
O=\left(|\lambda_{1}),\ |\lambda_{2}),\ |\lambda_{3}),\ |\lambda_{4})\right)=
\left(
\begin{array}{cccc}
1 & x_{2} & x_{3} & x_{4} \\
0 & y_{2} & y_{3} & y_{4} \\
0 & z_{2} & z_{3} & z_{4} \\
1 & w_{2} & 1 & 1
\end{array}
\right)
\end{equation}
we have $O\in GL(4;\fukuso)$ and
\[
O^{-1}=\frac{1}{|O|}
\left(
\begin{array}{cccc}
\widehat{O}_{11} & \widehat{O}_{12} & \widehat{O}_{13} & \widehat{O}_{14} \\
* & * & * & * \\
* & * & * & * \\
* & * & * & * 
\end{array}
\right)
\]
where $*$ denotes unnecessary terms in the following. 
Here, the cofactors are 
\[
\widehat{O}_{11}=
\left|
\begin{array}{ccc}
y_{2} & y_{3} & y_{4} \\
z_{2} & z_{3} & z_{4} \\
w_{2} & 1 & 1
\end{array}
\right|,
\widehat{O}_{12}=-
\left|
\begin{array}{ccc}
x_{2} & x_{3} & x_{4} \\
z_{2} & z_{3} & z_{4} \\
w_{2} & 1 & 1
\end{array}
\right|,
\widehat{O}_{13}=
\left|
\begin{array}{ccc}
x_{2} & x_{3} & x_{4} \\
y_{2} & y_{3} & y_{4} \\
w_{2} & 1 & 1
\end{array}
\right|,
\widehat{O}_{14}=-
\left|
\begin{array}{ccc}
x_{2} & x_{3} & x_{4} \\
y_{2} & y_{3} & y_{4} \\
z_{2} & z_{3} & z_{4}
\end{array}
\right|.
\]
We know that each term is very complicated. 
Note that
\begin{equation}
\label{eq:one relation}
|O|=\widehat{O}_{11}+\widehat{O}_{14} \ \Longrightarrow\ 
1=\frac{\widehat{O}_{11}}{|O|}+\frac{\widehat{O}_{14}}{|O|}.
\end{equation}

Now we are in a position to diagonalize $W$. Since
\[
W^{T}=OD_{W}O^{-1}
\]
with $D_{W}$ being the diagonal matrix 
\begin{equation}
\label{eq:diagonal part}
D_{W}=
\left(
\begin{array}{cccc}
0 &                  &                 &                 \\
   & \lambda_{2} &                 &                  \\
   &                 & \lambda_{3} &                  \\
   &                 &                 & \lambda_{4}
\end{array}
\right)
\end{equation}
we have
\begin{equation}
\label{eq:diagonal form}
W=(O^{T})^{-1}D_{W}O^{T}.
\end{equation}

Here, let us go back to the equation (\ref{eq:master equation 2}).
If we set
\[
(\hat{\rho}=)\Psi=
\left(
\begin{array}{c}
a         \\
b         \\
\bar{b} \\
d
\end{array}
\right)
\]
for simplicity, the equation (\ref{eq:master equation 2}) reads
\[
\frac{d}{dt}\Psi=W\Psi
\]
and the general solution is given by (\ref{eq:diagonal form})
\[
\Psi(t)=e^{tW}\Psi(0)=(O^{T})^{-1}e^{tD_{W}}O^{T}\Psi(0).
\]

Since we are interested in the final state $\Psi(\infty)$ we 
must look for the asymptotic limit 
$\lim_{t\rightarrow \infty}e^{tD_{W}}$.  From 
(\ref{eq:basic properties}) and (\ref{eq:diagonal part}) 
it is easy to see
\[
\lim_{t\rightarrow \infty}e^{tD_{W}}
=
\left(
\begin{array}{cccc}
1 & & & \\
& 0 & & \\
& & 0 & \\
& & & 0
\end{array}
\right)
=\kett{0} \braa{0},
\quad
\kett{0}\equiv 
\left(
\begin{array}{c}
1 \\
0 \\
0 \\
0
\end{array}
\right),
\]
so we obtain
\begin{equation}
\label{eq: limit form}
\Psi(\infty)
=(O^{T})^{-1}\kett{0}\braa{0}O^{T}\Psi(0)
=\frac{1}{|O|}
\left(
\begin{array}{cccc}
\widehat{O}_{11} & 0 & 0 & \widehat{O}_{11} \\
\widehat{O}_{12} & 0 & 0 & \widehat{O}_{12} \\
\widehat{O}_{13} & 0 & 0 & \widehat{O}_{13} \\
\widehat{O}_{14} & 0 & 0 & \widehat{O}_{14}
\end{array}
\right)\Psi(0).
\end{equation}

This equation gives
\[
\Psi(0)=
\left(
\begin{array}{c}
1 \\
0 \\
0 \\
0
\end{array}
\right)
\ \Longrightarrow \
\Psi(\infty)=\frac{1}{|O|}
\left(
\begin{array}{c}
\widehat{O}_{11} \\
\widehat{O}_{12} \\
\widehat{O}_{13} \\
\widehat{O}_{14}
\end{array}
\right)
\]
and it is equivalent to
\begin{equation}
\label{eq:result 1}
\rho_{0}(0)=\ket{0}\bra{0}=
\left(
\begin{array}{cc}
1 & 0 \\
0 & 0
\end{array}
\right)
\ \Longrightarrow \
\rho_{0}(\infty)=\frac{1}{|O|}
\left(
\begin{array}{cc}
\widehat{O}_{11} & \widehat{O}_{12} \\
\widehat{O}_{13} & \widehat{O}_{14}
\end{array}
\right).
\end{equation}

Similarly,
\[
\Psi(0)=
\left(
\begin{array}{c}
0 \\
0 \\
0 \\
1
\end{array}
\right)
\ \Longrightarrow \
\Psi(\infty)=\frac{1}{|O|}
\left(
\begin{array}{c}
\widehat{O}_{11} \\
\widehat{O}_{12} \\
\widehat{O}_{13} \\
\widehat{O}_{14}
\end{array}
\right)
\]
is equivalent to
\begin{equation}
\label{eq:result 2}
\rho_{1}(0)=\ket{1}\bra{1}=
\left(
\begin{array}{cc}
0 & 0 \\
0 & 1
\end{array}
\right)
\ \Longrightarrow \
\rho_{1}(\infty)=\frac{1}{|O|}
\left(
\begin{array}{cc}
\widehat{O}_{11} & \widehat{O}_{12} \\
\widehat{O}_{13} & \widehat{O}_{14}
\end{array}
\right).
\end{equation}

\vspace{3mm}
Let us state our result once more :
\begin{equation}
\label{eq:final result}
\rho_{0}(0)=\ket{0}\bra{0}, \quad
\rho_{1}(0)=\ket{1}\bra{1}
\ \Longrightarrow \ 
\rho_{0}(\infty)=\rho_{1}(\infty).
\end{equation}

\noindent
We would like to interpret the final density matrix as 
``classical one".

\vspace{3mm}
At the end of this section, let us present an important problem.

\noindent
{\bf Problem}\ \ Generalize the result to the case of 
$N$ level system of an atom.

For $N=3$ we conjecture that
\begin{eqnarray*}
\Psi(\infty)
&=&(O^{T})^{-1}\kett{0}\braa{0}O^{T}\Psi(0) \\
&=&\frac{1}{|O|}
\left(
\begin{array}{ccccccccc}
\widehat{O}_{11} & 0 & 0 & 0 & \widehat{O}_{11} & 0 & 0 & 0 & \widehat{O}_{11} \\
\widehat{O}_{12} & 0 & 0 & 0 & \widehat{O}_{12} & 0 & 0 & 0 & \widehat{O}_{12} \\
\widehat{O}_{13} & 0 & 0 & 0 & \widehat{O}_{13} & 0 & 0 & 0 & \widehat{O}_{13} \\
\widehat{O}_{14} & 0 & 0 & 0 & \widehat{O}_{14} & 0 & 0 & 0 & \widehat{O}_{14} \\
\widehat{O}_{15} & 0 & 0 & 0 & \widehat{O}_{15} & 0 & 0 & 0 & \widehat{O}_{15} \\
\widehat{O}_{16} & 0 & 0 & 0 & \widehat{O}_{16} & 0 & 0 & 0 & \widehat{O}_{16} \\
\widehat{O}_{17} & 0 & 0 & 0 & \widehat{O}_{17} & 0 & 0 & 0 & \widehat{O}_{17} \\
\widehat{O}_{18} & 0 & 0 & 0 & \widehat{O}_{18} & 0 & 0 & 0 & \widehat{O}_{18} \\
\widehat{O}_{19} & 0 & 0 & 0 & \widehat{O}_{19} & 0 & 0 & 0 & \widehat{O}_{19}
\end{array}
\right)\Psi(0)
\end{eqnarray*}
and
\begin{eqnarray*}
&&\rho_{0}(0)=\ket{0}\bra{0},\quad
\rho_{1}(0)=\ket{1}\bra{1},\quad 
\rho_{2}(0)=\ket{2}\bra{2}    \\
\Longrightarrow
&&\rho_{0}(\infty)=\rho_{1}(\infty)=\rho_{2}(\infty)
=\frac{1}{|O|}
\left(
\begin{array}{ccc}
\widehat{O}_{11} & \widehat{O}_{12} & \widehat{O}_{13} \\
\widehat{O}_{14} & \widehat{O}_{15} & \widehat{O}_{16} \\
\widehat{O}_{17} & \widehat{O}_{18} & \widehat{O}_{19}
\end{array}
\right)
\end{eqnarray*}
with some notations changed from $N=2$ to $N=3$.

\vspace{3mm}
We expect that young researchers will attack and 
solve the problem.

\section{Special Case}
The cubic equation is formally solved by 
the Cardano formula. However, in this case we cannot 
obtain a compact form of solutions \footnote{One can 
check this by MATHEMATICA}, so we assume
\begin{equation}
\label{eq:assumption}
E_{1}=E_{0}
\end{equation}
in this section. Then
\begin{equation}
\label{eq:special W}
W=
\left(
\begin{array}{cccc}
-\mu & i\bar{\gamma} & -i\gamma & \nu                        \\
i\gamma & -\frac{\mu+\nu}{2} & 0 & -i\gamma                 \\
-i\bar{\gamma} & 0 & -\frac{\mu+\nu}{2} & i\bar{\gamma}  \\
\mu & -i\bar{\gamma} & i\gamma & -\nu 
\end{array}
\right).
\end{equation}

The equation (\ref{eq:Cubic}) becomes
\[
\Lambda\left\{\Lambda^{2}+\frac{\mu+\nu}{2}\Lambda+4|\gamma|^{2}\right\}=0
\]
and the solutions are
\[
\Lambda_{0}=0,\quad
\Lambda_{\pm}=-\frac{\mu+\nu}{4}\pm
\frac{1}{2}\sqrt{\left(\frac{\mu+\nu}{2}\right)^{2}-16|\gamma|^{2}}.
\]

\noindent
Therefore, the eigenvalues of W (\ref{eq:special W}) are 
given by (from $\lambda=\Lambda-\frac{\mu+\nu}{2}$)
\begin{eqnarray}
\label{eq:eigenvalues}
&&\lambda_{1}=0,\ \ \lambda_{2}=-\frac{\mu+\nu}{2}, \nonumber \\
&&\lambda_{3}=-\frac{3}{4}(\mu+\nu)+\frac{1}{2}\sqrt{\left(\frac{\mu+\nu}{2}\right)^{2}-16|\gamma|^{2}},
\nonumber \\
&&\lambda_{4}=-\frac{3}{4}(\mu+\nu)-\frac{1}{2}\sqrt{\left(\frac{\mu+\nu}{2}\right)^{2}-16|\gamma|^{2}}.
\end{eqnarray}

For
\begin{equation}
\label{eq:special W^{T}}
W^{T}=
\left(
\begin{array}{cccc}
-\mu & i\gamma & -i\bar{\gamma} & \mu                        \\
i\bar{\gamma} & -\frac{\mu+\nu}{2} & 0 & -i\bar{\gamma}  \\
-i\gamma & 0 & -\frac{\mu+\nu}{2} & i\gamma                 \\
\nu & -i\gamma & i\bar{\gamma} & -\nu
\end{array}
\right)
\end{equation}
the corresponding eigenvectors are given by
\begin{eqnarray}
\label{eq:eigenvectors}
&&|\lambda_{1})=
\left(
\begin{array}{c}
1 \\
0 \\
0 \\
1
\end{array}
\right),
\quad
|\lambda_{2})=
\left(
\begin{array}{c}
0 \\
\bar{\gamma} \\
\gamma \\
0
\end{array}
\right),  \nonumber \\
&&|\lambda_{3})=
\left(
\begin{array}{c}
-1+\frac{2(\mu-\nu)}{\frac{\mu-7\nu}{4}+\frac{1}{2}\sqrt{\left(\frac{\mu+\nu}{2}\right)^{2}-16|\gamma|^{2}}}  \\
\\
2i\bar{\gamma}\ 
\frac{-1+\frac{\mu-\nu}{\frac{\mu-7\nu}{4}+\frac{1}{2}\sqrt{\left(\frac{\mu+\nu}{2}\right)^{2}-16|\gamma|^{2}}}}
       {-\frac{\mu+\nu}{4}+\frac{1}{2}\sqrt{\left(\frac{\mu+\nu}{2}\right)^{2}-16|\gamma|^{2}}}  \\
\\
-2i{\gamma}\ 
\frac{-1+\frac{\mu-\nu}{\frac{\mu-7\nu}{4}+\frac{1}{2}\sqrt{\left(\frac{\mu+\nu}{2}\right)^{2}-16|\gamma|^{2}}}}
       {-\frac{\mu+\nu}{4}+\frac{1}{2}\sqrt{\left(\frac{\mu+\nu}{2}\right)^{2}-16|\gamma|^{2}}}  \\
\\
1
\end{array}
\right),  \quad
|\lambda_{4})=
\left(
\begin{array}{c}
-1+\frac{2(\mu-\nu)}{\frac{\mu-7\nu}{4}-\frac{1}{2}\sqrt{\left(\frac{\mu+\nu}{2}\right)^{2}-16|\gamma|^{2}}}  \\
\\
2i\bar{\gamma}\ 
\frac{-1+\frac{\mu-\nu}{\frac{\mu-7\nu}{4}-\frac{1}{2}\sqrt{\left(\frac{\mu+\nu}{2}\right)^{2}-16|\gamma|^{2}}}}
       {-\frac{\mu+\nu}{4}-\frac{1}{2}\sqrt{\left(\frac{\mu+\nu}{2}\right)^{2}-16|\gamma|^{2}}}  \\
\\
-2i{\gamma}\ 
\frac{-1+\frac{\mu-\nu}{\frac{\mu-7\nu}{4}-\frac{1}{2}\sqrt{\left(\frac{\mu+\nu}{2}\right)^{2}-16|\gamma|^{2}}}}
       {-\frac{\mu+\nu}{4}-\frac{1}{2}\sqrt{\left(\frac{\mu+\nu}{2}\right)^{2}-16|\gamma|^{2}}}  \\
\\
1
\end{array}
\right). \nonumber \\
&&
\end{eqnarray}
Verification of the result is left to readers.

\section{Perturbation}
Since $\mu$ and $\nu$ in (\ref{eq:master equation 1}) 
are in general small compared to the terms in the Hamiltonian 
we can apply a perturbation method to the master equation 
(like \cite{KF1}) in order to obtain an approximate solution.

Let us decompose $W$ into two parts :
\begin{eqnarray*}
&&W=
\left(
\begin{array}{cccc}
-\mu & i\bar{\gamma} & -i\gamma & \nu                                          \\
i\gamma & i(E_{1}-E_{0})-\frac{\mu+\nu}{2} & 0 & -i\gamma                   \\
-i\bar{\gamma} & 0 & -i(E_{1}-E_{0})-\frac{\mu+\nu}{2} & i\bar{\gamma}  \\
\mu & -i\bar{\gamma} & i\gamma & -\nu 
\end{array}
\right) \\
&=&
\left(
\begin{array}{cccc}
0 & i\bar{\gamma} & -i\gamma & 0                          \\
i\gamma & i(E_{1}-E_{0}) & 0 & -i\gamma                   \\
-i\bar{\gamma} & 0 & -i(E_{1}-E_{0}) & i\bar{\gamma}  \\
0 & -i\bar{\gamma} & i\gamma & 0
\end{array}
\right)
+
\left(
\begin{array}{cccc}
-\mu & 0 & 0 & \nu                 \\
0 & -\frac{\mu+\nu}{2} & 0 & 0  \\
0 & 0 & -\frac{\mu+\nu}{2} & 0  \\
\mu & 0 & 0 & -\nu 
\end{array}
\right) \\
&\equiv& \widehat{H}+\widehat{D}.
\end{eqnarray*}
The general solution of (\ref{eq:master equation 2}) is given by
\begin{equation}
\label{eq:general solution}
\Psi(t)=e^{t\left(\widehat{H}+\widehat{D}\right)}\Psi(0).
\end{equation}
However, it is not easy to calculate the term 
$e^{t\left(\widehat{H}+\widehat{D}\right)}$ exactly, so 
we use a simple approximation
\[
e^{t\left(\widehat{H}+\widehat{D}\right)}=
e^{t\left(\widehat{D}+\widehat{H}\right)}\approx 
e^{t\widehat{D}}e^{t\widehat{H}}.
\]

In general, we must use the Zassenhaus formula (see 
for example \cite{Five}, \cite{CZ}). 

\vspace{3mm}\noindent
{\bf Zassenhaus Formula}\ \ For operators (or square matrices) 
$A$ and $B$ we have an expansion
\begin{equation}
\label{eq:Zassenhaus formula}
e^{t(A+B)}=
\cdots 
e^{-\frac{t^{3}}{6}\{2[[A,B],B]+[[A,B],A]\}}
e^{\frac{t^{2}}{2}[A,B]}
e^{tB}
e^{tA}.
\end{equation}
The formula is a bit different from that of \cite{CZ}. 

\noindent
From now on we discuss the approximate solution
\begin{equation}
\label{eq:general approximate solution}
\Psi(t)\approx e^{t\widehat{D}}e^{t\widehat{H}}\Psi(0).
\end{equation}

\vspace{3mm}
First, let us calculate $e^{t\widehat{D}}$. For the purpose 
we set
\[
K=
\left(
\begin{array}{cc}
-\mu & \nu \\
\mu & -\nu 
\end{array}
\right)
\]
and calculate $e^{tK}$. The eigenvalues of $K$ are 
$\{0,-(\mu+\nu)\}$ and corresponding eigenvectors (
not normalized) are
\[
0\longleftrightarrow 
\left(
\begin{array}{c}
\nu \\
\mu 
\end{array}
\right),\quad
-(\mu+\nu)\longleftrightarrow 
\left(
\begin{array}{c}
1  \\
-1
\end{array}
\right).
\]
If we define the matrix
\[
O=
\left(
\begin{array}{cc}
\nu & 1   \\
\mu & -1
\end{array}
\right)
\Longrightarrow 
O^{-1}=\frac{1}{\mu+\nu}
\left(
\begin{array}{cc}
1 & 1          \\
\mu & -\nu 
\end{array}
\right)
\]
then it is easy to see
\[
K=
O
\left(
\begin{array}{cc}
0 &                 \\
  & -(\mu+\nu)
\end{array}
\right)
O^{-1}
\]
and
\[
e^{tK}=
O
\left(
\begin{array}{cc}
1 &                         \\
   & e^{-t(\mu+\nu)}
\end{array}
\right)
O^{-1}
=
\frac{1}{\mu+\nu}
\left(
\begin{array}{cc}
\nu+\mu e^{-t(\mu+\nu)} & \nu-\nu e^{-t(\mu+\nu)}  \\
\mu-\mu e^{-t(\mu+\nu)} & \mu+\nu e^{-t(\mu+\nu)}
\end{array}
\right).
\]
Therefore, we have
\begin{equation}
\label{eq:E(tD)}
e^{t\widehat{D}}
=
\left(
\begin{array}{cccc}
\frac{\nu+\mu e^{-t(\mu+\nu)}}{\mu+\nu} & 0 & 0 & \frac{\nu-\nu e^{-t(\mu+\nu)}}{\mu+\nu}  \\
0 & e^{-t\frac{\mu+\nu}{2}} & 0 & 0                                                                                  \\
0 & 0 & e^{-t\frac{\mu+\nu}{2}} & 0                                                                                  \\
\frac{\mu-\mu e^{-t(\mu+\nu)}}{\mu+\nu} & 0 & 0 & \frac{\mu+\nu e^{-t(\mu+\nu)}}{\mu+\nu}
\end{array}
\right)
\approx
\frac{1}{\mu+\nu}
\left(
\begin{array}{cccc}
\nu & 0 & 0 & \nu   \\
0 & 0 & 0  & 0        \\
0 & 0 & 0  & 0        \\
\mu & 0 & 0 & \mu 
\end{array}
\right) 
\end{equation}
if $t$ is large enough ($t\gg 1/(\mu+\nu)$).

Next, let us calculate $e^{t\widehat{H}}$. Since we need 
some properties of tensor products in the following see 
for example \cite{Five}. We can express $\widehat{H}$ 
as
\[
\widehat{H}=-i\left(H\otimes 1_{2}-1_{2}\otimes H^{T}\right).
\]
In fact,
\begin{eqnarray*}
\widehat{H}
&=&-i
\left\{
\left(
\begin{array}{cc}
E_{0}    & \gamma       \\
\bar{\gamma } & E_{1}
\end{array}
\right)
\otimes
\left(
\begin{array}{cc}
1 & 0 \\
0 & 1
\end{array}
\right)
-
\left(
\begin{array}{cc}
1 & 0 \\
0 & 1
\end{array}
\right)
\otimes
\left(
\begin{array}{cc}
E_{0}    & \bar{\gamma} \\
\gamma & E_{1}
\end{array}
\right)
\right\} \\
&=&-i
\left\{
\left(
\begin{array}{cccc}
E_{0} & 0 & \gamma  & 0        \\
0 & E_{0} & 0 & \gamma         \\
\bar{\gamma} &0 & E_{1} & 0  \\
0 & \bar{\gamma} & 0 & E_{1}
\end{array}
\right)
-
\left(
\begin{array}{cccc}
E_{0} & \bar{\gamma} & 0 & 0 \\
\gamma & E_{1} & 0 & 0         \\
0 & 0 & E_{0} & \bar{\gamma} \\
0 & 0 & \gamma & E_{1}
\end{array}
\right)
\right\} \\
&=&-i
\left(
\begin{array}{cccc}
0 & -\bar{\gamma} & \gamma & 0                    \\
-\gamma & -(E_{1}-E_{0}) & 0 & \gamma            \\
\bar{\gamma} & 0 & E_{1}-E_{0} & -\bar{\gamma} \\
0 & \bar{\gamma} & -\gamma & 0
\end{array}
\right).
\end{eqnarray*}

It is well--known that
\[
e^{t\widehat{H}}
=
e^{-it\left(H\otimes 1_{2}-1_{2}\otimes H^{T}\right)}
=
e^{-it H\otimes 1_{2}}e^{it 1_{2}\otimes H^{T}}
=
\left(e^{-itH}\otimes 1_{2}\right)
\left(1_{2}\otimes e^{itH^{T}}\right)
=
e^{-itH}\otimes e^{itH^{T}},
\]
so we must calculate
\[
e^{-itH}=\exp
\left\{-it
\left(
\begin{array}{cc}
E_{0} & \gamma        \\
\bar{\gamma} & E_{1}
\end{array}
\right)
\right\}.
\]
Since $H$ in (\ref{eq:interaction matrix}) is expressed as
\begin{eqnarray*}
\left(
\begin{array}{cc}
E_{0} & \gamma        \\
\bar{\gamma} & E_{1}
\end{array}
\right)
&=&
\left(
\begin{array}{cc}
\frac{E_{0}+E_{1}}{2} &  \\
 & \frac{E_{0}+E_{1}}{2} 
\end{array}
\right)
+
\left(
\begin{array}{cc}
-\frac{E_{1}-E_{0}}{2} & \gamma       \\
\bar{\gamma} & \frac{E_{1}-E_{0}}{2} 
\end{array}
\right) \\
&\equiv& 
\Delta_{+}{1}_{2}+
\left(
\begin{array}{cc}
-\Delta_{-} & \gamma       \\
\bar{\gamma} & \Delta_{-}
\end{array}
\right)
\quad \mbox{where}\quad \Delta_{\pm}=\frac{E_{1}\pm E_{0}}{2}
\end{eqnarray*}
the calculation is reduced to
\[
e^{-itH}=e^{-it\Delta_{+}}
\exp
\left\{-it
\left(
\begin{array}{cc}
-\Delta_{-} & \gamma       \\
\bar{\gamma} & \Delta_{-}
\end{array}
\right)
\right\}.
\]
This exponential is well--known, see for example \cite{Five}. 
That is,
\begin{equation}
\label{eq:calculation-1}
\exp
\left\{-it
\left(
\begin{array}{cc}
-\Delta_{-} & \gamma       \\
\bar{\gamma} & \Delta_{-}
\end{array}
\right)
\right\}
=
\left(
\begin{array}{cc}
a_{11} & a_{12} \\
a_{21} & a_{22}
\end{array}
\right)
\end{equation}
where
\begin{eqnarray}
\label{eq:components}
a_{11}&=&\cos(t\sqrt{\Delta_{-}^{2}+|\gamma|^{2}})+
i\frac{\sin(t\sqrt{\Delta_{-}^{2}+|\gamma|^{2}})}{\sqrt{\Delta_{-}^{2}+|\gamma|^{2}}}\Delta_{-}, 
\nonumber \\
a_{12}&=&-i\frac{\sin(t\sqrt{\Delta_{-}^{2}+|\gamma|^{2}})}{\sqrt{\Delta_{-}^{2}+|\gamma|^{2}}}\gamma, 
\nonumber \\
a_{21}&=&-i\frac{\sin(t\sqrt{\Delta_{-}^{2}+|\gamma|^{2}})}{\sqrt{\Delta_{-}^{2}+|\gamma|^{2}}}\bar{\gamma}, 
\nonumber \\
a_{22}&=&\cos(t\sqrt{\Delta_{-}^{2}+|\gamma|^{2}})-
i\frac{\sin(t\sqrt{\Delta_{-}^{2}+|\gamma|^{2}})}{\sqrt{\Delta_{-}^{2}+|\gamma|^{2}}}\Delta_{-}.
\end{eqnarray}
Similarly, we obtain
\[
e^{itH^{T}}=e^{it\Delta_{+}}
\left(
\begin{array}{cc}
a_{22} & -a_{21} \\
-a_{12} & a_{11}
\end{array}
\right).
\]

Therefore, we arrive at
\begin{eqnarray}
\label{eq:E(tH)}
e^{-itH}\otimes e^{itH^{T}}
&=&
\left(
\begin{array}{cc}
a_{11} & a_{12} \\
a_{21} & a_{22}
\end{array}
\right)
\otimes
\left(
\begin{array}{cc}
a_{22} & -a_{21} \\
-a_{12} & a_{11}
\end{array}
\right) \nonumber \\
&=&
\left(
\begin{array}{cccc}
a_{11}a_{22} & -a_{11}a_{12} & a_{12}a_{22} & -a_{12}a_{21} \\
* & * & * & *                                                         \\
* & * & * & *                                                         \\
-a_{12}a_{21} & a_{11}a_{21} & -a_{12}a_{22} & a_{11}a_{22} 
\end{array}
\right)  \nonumber \\
&\equiv&
\left(
\begin{array}{cccc}
c_{11} & c_{12} & c_{13} & c_{14}  \\
* & * & * & *                         \\
* & * & * & *                         \\
c_{41} & c_{42} & c_{43} & c_{44}
\end{array}
\right)
\end{eqnarray}
where $*$'s in the matrix are elements not used in later discussion.

From (\ref{eq:E(tH)}) and (\ref{eq:components}) 
it is easy to see
\begin{equation}
\label{eq: important relation}
c_{11}+c_{41}=1,\quad
c_{12}+c_{42}=0,\quad
c_{13}+c_{43}=0,\quad
c_{14}+c_{44}=1.
\end{equation}

Therefore, from (\ref{eq:general approximate solution}), 
(\ref{eq:E(tD)}), (\ref{eq:E(tH)}) and (\ref{eq: important relation}) 
we obtain
\begin{eqnarray}
\label{eq:Fujii}
\Psi(t)
&\approx& 
\frac{1}{\mu+\nu}
\left(
\begin{array}{cccc}
\nu & 0 & 0 & \nu   \\
0 & 0 & 0  & 0        \\
0 & 0 & 0  & 0        \\
\mu & 0 & 0 & \mu 
\end{array}
\right)
\left(
\begin{array}{cccc}
c_{11} & c_{12} & c_{13} & c_{14}  \\
* & * & * & *                         \\
* & * & * & *                         \\
c_{41} & c_{42} & c_{43} & c_{44}
\end{array}
\right)
\Psi(0)
\nonumber \\
&=& 
\frac{1}{\mu+\nu}
\left(
\begin{array}{cccc}
\nu & 0 & 0 & \nu   \\
0 & 0 & 0  & 0        \\
0 & 0 & 0  & 0        \\
\mu & 0 & 0 & \mu 
\end{array}
\right)
\Psi(0)
\end{eqnarray}
for $t\gg 1/(\mu+\nu)$.

From (\ref{eq:density matrix})
\[
\rho(t)=
\left(
\begin{array}{cc}
a(t)         & b(t)  \\
\bar{b}(t) & d(t)
\end{array}
\right),
\quad
\rho(0)=
\left(
\begin{array}{cc}
a(0)         & b(0)  \\
\bar{b}(0) & d(0)
\end{array}
\right)
\]
we have
\begin{eqnarray}
\rho(\infty)
&=&
\frac{1}{\mu+\nu}
\left(
\begin{array}{cc}
\nu\left(a(0)+d(0)\right) & 0  \\
0 & \mu\left(a(0)+d(0)\right) 
\end{array}
\right)  
=
\frac{1}{\mu+\nu}
\left(
\begin{array}{cc}
\nu & 0  \\
0 & \mu
\end{array}
\right)  \nonumber \\
&=&
\frac{\nu}{\mu+\nu}\ket{0}\bra{0}+\frac{\mu}{\mu+\nu}\ket{1}\bra{1}
\end{eqnarray}
because $\mbox{tr}\rho(0)=a(0)+d(0)=1$

\vspace{3mm}
We believe that the result in this section is deeply 
related to the proof of the Copenhagen interpretation, 
see \cite{KF1}.

\section{Concluding Remarks}
In this paper we have derived the solutions to the master 
equation of the two level system of an atom under decoherence. 
How do we understand the result from the physical point of view ? 
We would like to interpret the final density matrix as a  
representation of some classical state. 

In general, to solve a master equation exactly is very hard, 
so we are usually satisfied by solving it approximately. 
For example, see \cite{KF1} and \cite{KF3}. 
As far as we know our result is the finest one up to the present.

We want to apply the results in the paper to our method 
of Quantum Computation based on Cavity QED, see \cite{FHKW1} 
and \cite{FHKW2}. In the quantum computation we must 
take {\bf decoherence time} into consideration, which is 
an essential point. 
Some results will be reported in the near future.

In standard textbooks of QM  decoherence theory is usually not 
contained, so it may be hard for beginners (young students) 
to understand. 
For example a book \cite{BP} or a recent review paper \cite{KH} 
would be very helpful for beginners.

\vspace{5mm}\noindent 
{\it Acknowledgments}\ \ 
The author wishes to thank Ryu Sasaki for useful suggestions and comments.


\end{document}

%% file: eigenvalues.tex
\unitlength 0.1in
\begin{picture}( 33.2500, 17.9500)( 12.0000,-21.7500)
%
\special{pn 8}%
\special{pa 1726 1176}%
\special{pa 2716 1176}%
\special{fp}%
%
\special{pn 8}%
\special{pa 1726 1776}%
\special{pa 2716 1776}%
\special{fp}%
%
\special{pn 8}%
\special{pa 3536 776}%
\special{pa 4526 776}%
\special{fp}%
%
\special{pn 8}%
\special{pa 3526 2176}%
\special{pa 4516 2176}%
\special{fp}%
%
\special{pn 8}%
\special{pa 2776 1116}%
\special{pa 3446 806}%
\special{dt 0.045}%
\special{sh 1}%
\special{pa 3446 806}%
\special{pa 3376 816}%
\special{pa 3398 828}%
\special{pa 3394 852}%
\special{pa 3446 806}%
\special{fp}%
%
\special{pn 8}%
\special{pa 2756 1846}%
\special{pa 3426 2156}%
\special{dt 0.045}%
\special{sh 1}%
\special{pa 3426 2156}%
\special{pa 3374 2110}%
\special{pa 3378 2134}%
\special{pa 3356 2146}%
\special{pa 3426 2156}%
\special{fp}%
\put(15.1500,-11.7500){\makebox(0,0){$E_{1}$}}%
\put(15.3500,-17.7500){\makebox(0,0){$E_{0}$}}%
\put(47.6500,-7.7500){\makebox(0,0){$\lambda_{+}$}}%
\put(47.7500,-21.7500){\makebox(0,0){$\lambda_{-}$}}%
\put(21.5500,-4.7500){\makebox(0,0){$\gamma=0$}}%
\put(39.8500,-4.6500){\makebox(0,0){$\gamma\ne 0$}}%
\end{picture}%

%% file: ExactSol_MasterEq_II.bbl
\begin{thebibliography}{99}
%
\bibitem{PD}P. Dirac : 
\newblock {\bf The Principles of Quantum Mechanics}, 
\newblock Fourth Edition, Oxford University Press, 1958.
%
\bibitem{HG}H. S. Green : 
\newblock {\bf Matrix Mechanics}, 
\newblock P. Noordhoff Ltd, Groningen, 1965.
%
\bibitem{AP}Asher Peres : 
\newblock {\bf Quantum Theory : Concepts and Methods}, 
\newblock Kluwer Academic Publishers, 1995.
%
\bibitem{AH}Akio Hosoya : 
\newblock {\bf Lectures on Quantum Computation} (in Japanese), 
\newblock SGC Library 4, Saiensu-sha Co., Ltd. Publishers (Tokyo), 1999.
%
\bibitem{WZ}W. H. Zurek :
\newblock Decoherence and the transition from quantum to classical, 
\newblock Physics Today, {\bf 44} (1991), 36-44.
%
\bibitem{WS}W. P. Schleich : 
\newblock {\bf Quantum Optics in Phase Space},
\newblock WILEY--VCH, Berlin, 2001.
%
\bibitem{Five}K. Fujii and et al : 
\newblock {\bf Treasure Box of Mathematical Sciences} (in Japanese),
\newblock Yuseisha, Tokyo, 2010. \\
\newblock I expect that the book will be translated into English.
%
\bibitem{TM}Toshihide Maskawa :
\newblock {\bf Yet, Another Introduction to Elementary Particle Theory} (in Japanese), 
\newblock Maruzen Ltd, Tokyo, 1998.
%
\bibitem{Lind}G. Lindblad : 
\newblock On the generator of quantum dynamical semigroups, 
\newblock Commun. Math. Phys, {\bf 48} (1976), 119.
%
\bibitem{GKS} V. Gorini, A. Kossakowski and E. C. G. Sudarshan : 
\newblock Completely positive dynamical semigroups of N--level systems, 
\newblock J. Math. Phys, {\bf 17} (1976), 821.
%
\bibitem{BP}H. -P. Breuer and F. Petruccione : 
\newblock {\bf The theory of open quantum systems}, 
\newblock Oxford University Press, New York, 2002.
%
\bibitem{KF2}K. Fujii : 
\newblock Quantum Damped Harmonic Oscillator, 
\newblock {\bf Chapter 7} of ``Advances in Quantum Mechanics",
Paul Bracken (Ed.), ISBN 978-953-51-1089-7, InTech, 
\newblock arXiv:1209.1437 [quant-ph].
%
\bibitem{KF4}K. Fujii :
\newblock A modern introduction to Cardano and Ferrari formulas 
in the algebraic equations, 
\newblock Far East Journal of Mathematical Education, {\bf 10}(2013), 175-189, 
\newblock quant-ph/0311102.
%
\bibitem{KF1}K. Fujii :
\newblock ``Proof " of the Copenhagen Interpretation, 
\newblock arXiv:1304.1591 [quant-ph].
%
\bibitem{CZ}C. Zachos :
\newblock Crib Notes on Campbell-Baker-Hausdorff expansions, 
\newblock unpublished, 1999,
\newblock see\ http://www.hep.anl.gov/czachos/index.html.
%
\bibitem{KF3}K. Fujii :
\newblock Superluminal Group Velocity of Neutrinos : Review, Development 
and Problems, 
\newblock Int. J. Geom. Methods Mod. Phys, {\bf 10} (2013), 1250083 (19 pages), 
\newblock arXiv:1203.6425 [physics].
%
\bibitem{FHKW1}K. Fujii, K. Higashida, R. Kato and Y. Wada : 
\newblock Cavity QED and Quantum Computation in the Weak Coupling Regime, 
\newblock J. Opt. B : Quantum and Semiclass. Opt, {\bf 6} (2004), 502, 
\newblock quant-ph/0407014. 
%
\bibitem{FHKW2}K. Fujii, K. Higashida, R. Kato and Y. Wada : 
\newblock Cavity QED and Quantum Computation in the Weak Coupling Regime II : 
Complete Construction of the Controlled--Controlled NOT Gate, 
\newblock Trends in Quantum Computing Research, Susan Shannon (Ed.), 
{\bf Chapter 8}, Nova Science Publishers, 2006 and 
Computer Science and Quantum Computing, James E. Stones (Ed.), 
{\bf Chapter 1}, Nova Science Publishers, 2007, 
\newblock quant-ph/0501046. 
%
\bibitem{KH}K. Hornberger :
\newblock Introduction to Decoherence Theory, 
\newblock in ``Theoretical Foundations of Quantum Information", 
Lecture Notes in Physics, {\bf 768} (2009), 221-276, Springer, Berlin, 
\newblock quant-ph/061211.
%
\end{thebibliography}
